\documentclass[twocolumn]{revtex4}
\usepackage{times}
\usepackage{graphicx}
\def\lesssim{\mathrel{\mathpalette\vereq<}}
\def\gtrsim{\mathrel{\mathpalette\vereq>}}

\begin{document}
\title{Neutrinoless Double Beta Decay in Light of SNO Salt Data}
\author{Hitoshi Murayama\footnote{On leave of absence from Department
of Physics, University of California, Berkeley, California 94720, USA
} and Carlos Pe\~na-Garay}
\affiliation{School of Natural Sciences, Institute for Advanced Study,
Einstein Dr, Princeton, NJ 08540, USA}

\date{\today}
\begin{abstract}
  In the SNO data from its salt run, probably the most significant
  result is the consistency with the previous results without assuming
  the $^8$B energy spectrum.  In addition, they have excluded the
  maximal mixing at a very high confidence level.  This has an
  important implication on the double beta decay experiments.  For the
  inverted or degenerate mass spectrum, we find $|\langle
  m_\nu\rangle_{ee}| > 0.013$~eV at 95\% CL, and the next generation
  experiments can discriminate Majorana and Dirac neutrinos if the
  inverted or degenerate mass spectrum will be confirmed by the
  improvements in cosmology, tritium data beta decay, or long-baseline
  oscillation experiments.
\end{abstract}
\pacs{Who cares?}
\maketitle


In the past five years, there had been an amazing progress in neutrino
physics.  The atmospheric neutrinos showed a large up-down asymmetry
in the SuperKamiokande (SK) experiment which came as the first
significant evidence for a finite neutrino mass \cite{SuperK} and
hence the incompleteness of the Standard Model of particle physics.
SuperKamiokande also improved the accuracy in solar neutrino studies
greatly using the elastic scattering (ES) process.  The Sudbury
Neutrino Observatory (SNO) experiment has studied the charged-current
(CC) and neutral-current (NC) process in addition to the ES process,
and has shown that the solar neutrinos change their flavors from the
electron type to other active types (muon and tau neutrinos)
\cite{SNO}.  Finally, the KamLAND reactor anti-neutrino oscillation
experiments reported a significant deficit in reactor anti-neutrino
flux over approximately 180~km of propagation \cite{KamLAND}.  Further
combined with the pioneering Homestake experiment \cite{Homestake} and
Gallium-based experiments \cite{Gallium}, the decades-long solar
neutrino problem \cite{solarproblem} appears solved.  The so-called
Large Mixing Angle (LMA) solution \cite{review}, where the electron
neutrinos produced at Sun's core propagate adiabatically to a heavier
mass eigenstate due to the matter effect \cite{MSW}, is the only
viable explanation of the data.

On September 7, 2003, SNO published the result from their salt run
with an enhanced sensitivity to the NC process \cite{SNOnew} .  Most
importantly, the new result agrees well with previous results,
confirming the LMA solution to the solar neutrino problem.  In
addition, they have reported a much better determination of the mixing
angle $\theta_{12}$, which excludes the maximal mixing
$\theta_{12}=\pi/4$ at a very high significance: 5.4 sigma.

The exclusion of the maximal mixing has an important impact on another
crucial question in neutrino physics: Is neutrino its own
anti-particle?  If yes, neutrinos are Majorana fermions; if not,
Dirac.  This question is even deeper than it sounds.  For instance, if
neutrinos and anti-neutrinos are identical, there could have been a
process in Early Universe that affected the balance between particles
and anti-particles, leading to the matter anti-matter asymmetry we
need to exist.  In fact, so-called leptogenesis models directly link
the Majorana nature of neutrinos to the observed baryon asymmetry
\cite{leptogenesis} \footnote{It is possible, however, that the
  leptogenesis occurs even with Dirac neutrinos \cite{Dirac}.}.

This question can in principle be resolved if a neutrinoless double
beta decay is observed.  Because such a phenomenon will violate the
lepton number by two units, it cannot be caused if the neutrino is
different from the anti-neutrino \footnote{Other new physics beyond
  the Standard Model, {\it e.g.}\/ $R$-parity violating supersymmetry,
  Majoron, doubly charged Higgs, can also cause neutrinoless
  double-beta decay (see {\it e.g.}\/, \cite{Fukugita:en} and
  references therein).  However, such models induce Majorana mass for
  neutrinos from radiative corrections as well.}.  Many experimental
proposals exist that will increase the sensitivity to such a
phenomenon dramatically over the next ten years \cite{Gratta}.  The
crucial question is if a negative result from such experiments can
lead to a definitive statement about the nature of neutrinos.  In
particular, the matrix element of neutrinoless double beta decay is
proportional to the effective electron-neutrino mass \cite{Vogel}
\begin{equation}
   \langle m_\nu\rangle_{ee} = m_1 U_{e1}^2 + m_2 U_{e2}^2 + m_3 U_{e3}^2,
\end{equation}
which may have cancellation among three terms that makes it difficult
to assess the result of a negative search.  However, the exclusion of
the maximal mixing in $\theta_{12}$ actually helps to eliminate such
an unfortunate situation.  Note that the proposed experiments are
aiming at the sensitivity reaching $|\langle m_\nu\rangle_{ee}| \sim
0.01$~eV \cite{Gratta}.

\begin{figure}
  \includegraphics[width=0.9\columnwidth]{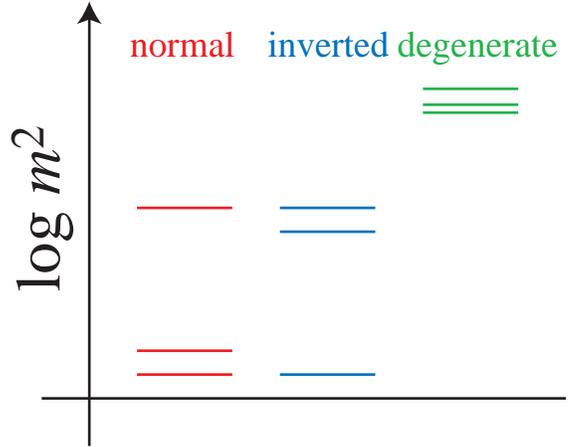}
  \caption{\label{fig:spectra}Three possible mass spectra of
    neutrinos.  The wider splitting is $\Delta m^2_{\rm atm} \simeq 2
    \times 10^{-3}$~eV$^2$, while the smaller one is $\Delta m^2_{\rm
      solar} \simeq 7 \times 10^{-5}$~eV$^2$.}
\end{figure}

\begin{figure*}

  \includegraphics[width=1.8\columnwidth]{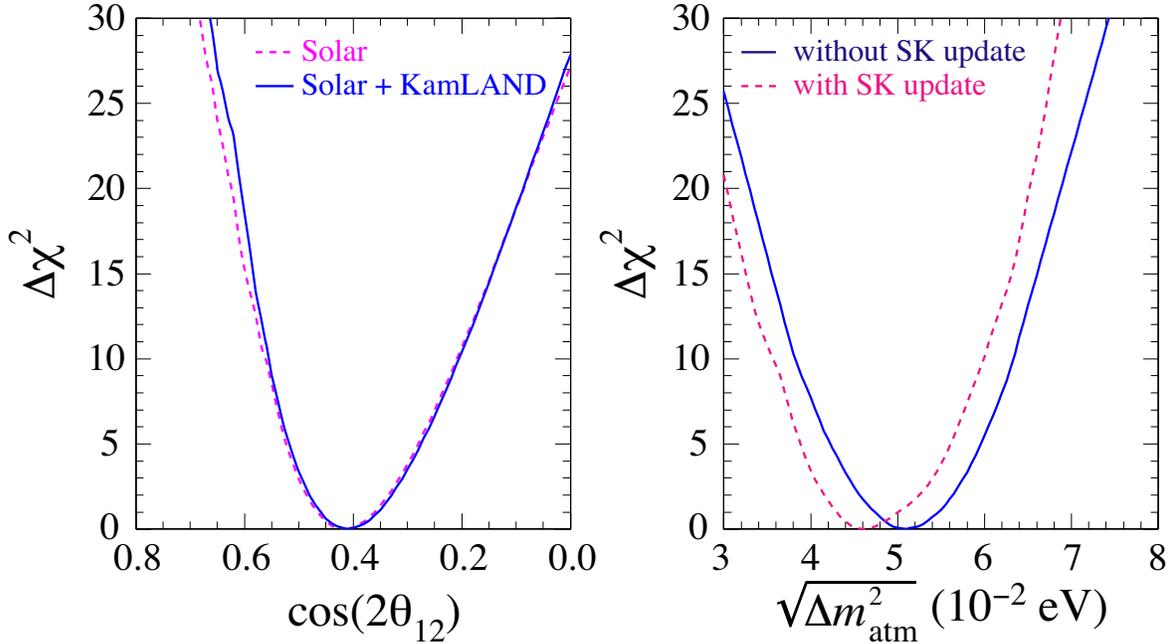}
\caption{\label{fig:sun_atm} Left: $\Delta \chi^2$ vs $\cos 2
  \theta_{12}$ with all solar neutrino data with and without KamLAND,
  marginalized on $\Delta m^2_{12}$. Right: $\Delta \chi^2$ vs
  $\sqrt{\Delta m^2_{\rm atm}}$ with SuperKamiokande without the
  update \protect\cite{Gonzalez-Garcia:2003qf} or with the update
  \cite{Kajita}, marginalized on $\sin^2 2\theta_{23}$ and combined
  with K2K.}
\end{figure*}

Within three generations of neutrinos and given all neutrino
oscillation data \footnote{The positive evidence for neutrino
  oscillation from the LSND experiment \cite{LSND} does not fit into
  the standard three-generation framework.  We ignore this evidence in
  this letter.}, there are three possible mass spectra: degenerate,
normal hierarchy and inverted hierarchy (see Fig.~\ref{fig:spectra})
\footnote{The degenerate spectrum can further be either of normal or
  inverted hierarchy.}.  Given that the third mixing angle
$\theta_{13} = \arcsin |U_{e3}|$ is known to be small from the CHOOZ
limit \cite{CHOOZ}, one can obtain a lower bound on the effective
electron-neutrino mass.  For the degenerate spectrum of the nearly
common mass $m$, we can ignore $m_3 U_{e3}^2$ relative to two other
terms, and find
\begin{eqnarray}
   & & |\langle m_\nu\rangle_{ee}| \simeq |m U_{e1}^2 + m U_{e2}^2|
   \nonumber \\
   & & \geq m (|U_{e1}|^2 - |U_{e2}|^2) =  m \cos 2\theta_{12}.
\end{eqnarray}
For the inverted hierarchy, $m_1 \simeq m_2 \geq \sqrt{\Delta
  m^2_{\rm atm}}$, and we can again ignore $m_3 U_{e3}^2$ relative to
two other terms.  Therefore,
\begin{eqnarray}
   & & |\langle m_\nu\rangle_{ee}| 
   \geq \sqrt{\Delta m^2_{\rm atm}} |U_{e1}^2 + U_{e2}^2|
   \nonumber \\
   & & \geq \sqrt{\Delta m^2_{\rm atm}} (|U_{e1}|^2 - |U_{e2}|^2) 
   =  \sqrt{\Delta m^2_{\rm atm}} \cos 2\theta_{12}.
   \label{eq:main}
\end{eqnarray}
Note that the bound for the inverted hierarchy is weaker than that for
the degenerate spectrum by definition, because the degeneracy requires
$m \gtrsim \sqrt{\Delta m^2_{\rm atm}}$.  Therefore,
Eq.~(\ref{eq:main}) is our master equation for most of our
discussions.  

Unfortunately, for the normal hierarchy, one cannot obtain a similar
rigorous lower limit.  On the other hand, the improvement in the
cosmological data \cite{Hu:1997mj} and the KATRIN experiment on the
end point of the trium beta decay \cite{KATRIN} may positively
establish the degenerate spectrum, or the long baseline neutrino
oscillation experiments may positively establish the inverted
hierarchy \cite{LBL}.  If either of them happens, and if the
neutrinoless double beta won't be seen within these bounds, the
neutrinos will be found to be Dirac particles \footnote{This statement
  assumes that there are only three light neutrinos mass eigenstates.
  HM thanks Xerxes Tata on this point.  See \cite{Beacom:2003eu} for a
  related discussion.}. 

$\cos 2\theta_{12}$ now has a robust lower bound given the new SNO
result.  To best of our knowledge, it was pointed out first in
\cite{Bilenky:1996cb} that the less than maximal mixing leads to a
lower bound on $|\langle m_\nu\rangle_{ee}|$ for the degenerate and
inverted spectra.  More recent papers \cite{recent} studied the bound
quantitatively before the recent SNO result when the lower bound was
not quite robust, because the exclusion of the maximal mixing was
reported at different confidence levels among different analyses and
depended crucially on Homestake data \cite{Homestake}.

There are obviously two main ingredients in the lower bound.  One is
$\Delta m^2_{\rm atm}$ from SuperKamiokande experiment which had
recently been updated \cite{Hayato}, and the other is $\theta_{12}$
from the solar neutrino data which includes the recent SNO result.
The last ingredient is $\theta_{13}$ which we assume to be zero
throughout our discussions.  We will come back to the little effect of
non-vanishing $\theta_{13}$ at the end of the letter.

First on $\Delta m^2_{\rm atm}$.  The analysis of the atmospheric (SK)
and accelerator (K2K) data was done in the general case of 3$\nu$
oscillations in \cite{Gonzalez-Garcia:2003qf}, and we show the
marginalized $\Delta \chi^2$ as a function of $\sqrt{\Delta m^2_{\rm
    atm}}$ in the right pane of Fig.~\ref{fig:sun_atm}.  The
constraint $\theta_{13}=0$ does not modify the shape of these
functions. This analysis uses the data available before updates this
summer \cite{Hayato}.  The SK preliminary analysis of atmospheric data
show a shift of the allowed region to lower $\Delta m^2_{\rm atm}$,
due to several improvements in their analysis: new neutrino flux with
updated primary cosmic ray flux, hadron interaction model and
calculation methods (3D), and improved neutrino interactions, detector
simulation and event reconstruction.  We included the SK update
\cite{Kajita}. 

Second on $\theta_{12}$.  The analysis of solar and reactor data is
done as described in \cite{Gonzalez-Garcia:2003qf}, except that
$\theta_{13}$ is set to zero, the Gallium rate is updated
\cite{Gallium} and the latest SNO data (NC, CC and ES measured in
phase-II \cite{SNOnew}) is included \cite{preparation}.  The $\Delta
\chi^2$ is shown as a function of $\cos 2\theta_{12}$ in the right
pane of Fig.~\ref{fig:sun_atm}.

\begin{figure}
\includegraphics[width=0.9\columnwidth]{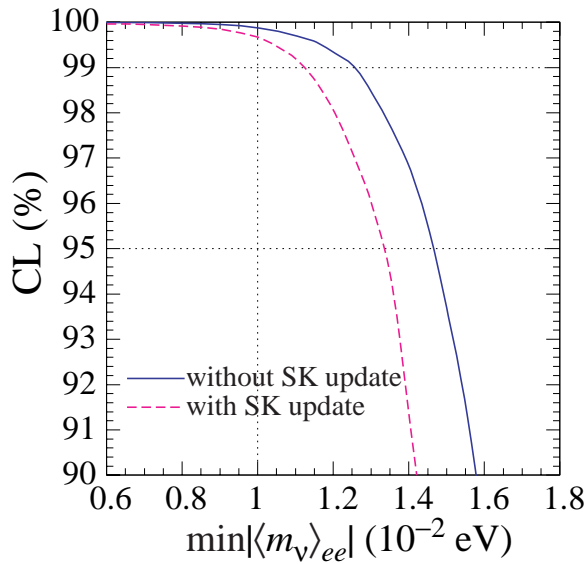}
\caption{\label{fig:cl} Lower bound on $|\langle m_\nu\rangle_{ee}|$ vs
  confidence levels for the inverted hierarchy spectrum.  Note that
  what is shown here is $\sqrt{\Delta m^2_{\rm atm}} \cos
  2\theta_{12}$, which is the minimum value of $|\langle
  m_\nu\rangle_{ee}|$ allowing the maximum cancellation between $m_1$
  and $m_2$.  The solid line is based on the atmospheric neutrino data
  before the update, while the dashed line with the update.  }
\end{figure}

Combining $\Delta m^2_{\rm atm}$ and $\theta_{12}$ discussed above, we
obtain the final result on the effective electron-neutrino mass.  The
lower bounds are shown at different confidence levels in
Fig.~\ref{fig:cl}.  One can see that 
\begin{equation}
   |\langle m_\nu\rangle_{ee}| > 0.013~(011)~\mbox{eV} \qquad \mbox{95\%
   CL (99\% CL)}.
\end{equation}
The result is quite robust in the sense that one has to require an
extremely high confidence level (99.7\%) to bring $|\langle m_\nu
\rangle_{ee}|$ below 0.01~eV.  Recall that the proposed experiments are aiming
at the sensitivity reaching $|\langle m_\nu\rangle_{ee}| \sim 0.01$~eV
\cite{Gratta}.

In all of the above discussions, we ignored $\theta_{13}$.  First of
all, $\theta_{13}$ is small due to the limit from CHOOZ reactor
experiment \cite{CHOOZ}.  Even setting the CHOOZ limit aside, it
is well-known, however, that $\theta_{13}$ has very little effect on
the determination of $\Delta m^2_{\rm atm}$
\cite{Gonzalez-Garcia:2003qf}, and also can only decrease the
preferred values of $\theta_{12}$
\cite{Fogli:1999zg,Gonzalez-Garcia:2000sq}.  Therefore, the impact of
a non-vanishing $\theta_{13}$ on $\Delta m^2_{\rm atm}$ and
$\theta_{12}$ can only strengthens our result.

One may also worry about corrections to the approximate formula
Eq.~(\ref{eq:main}) due to $\Delta m^2_{\rm solar}$ and $\theta_{13}$.
To minimize $|\langle m_\nu\rangle_{ee}|$, we can study the case where
both $U_{e2}^2$ and $U_{e3}^2$ have the opposite sign from $U_{e1}^2$,
giving
\begin{eqnarray}
  \lefteqn{
    |\langle m_\nu\rangle_{ee}| = |(m_1-m_3) c_{13}^2 \cos 2\theta_{12}
  }
  \nonumber \\
  & &
  - (m_2-m_1) c_{13}^2 s_{12}^2 
  + m_3 (c_{13}^2 \cos 2\theta_{12} - s_{13}^2)|.
\end{eqnarray}
In the limit $\Delta m_{\rm solar}^2=0$ ($m_2=m_1$) and
$\theta_{13}=0$, it reduces to Eq.~(\ref{eq:main}) due to the first
term above.  The suppression factor due to $c_{13}^2$ is at most 4.4\%
(95\% CL) thanks to the CHOOZ limit.  The second term does not vanish
due to $\Delta m_{\rm solar}^2 \neq 0$, and gives a correction at most
of
\begin{equation}
  \frac{\Delta m^2_{\rm solar}}{2\Delta m^2_{\rm atm}}
  \frac{s_{12}^2}{\cos 2\theta_{12}} \lesssim 3\% \quad \mbox{(95\% CL)}.
\end{equation}
Finally, the last term cannot be negative given the CHOOZ limit and
only strengthens our limit.  Overall, our lower bound can change at
most by 8\%.

The bound on $|\langle m_\nu\rangle_{ee}|$ is expected to improve
further as more data will become available.  As for long-baseline
(LBL) accelerator-based neutrino oscillation experiments, K2K will
double the data set, while MINOS, ICARUS, and OPERA are expected to
come online around 2005.  If approved, the neutrino beam from J-PARC
will be available around 2007.  They will improve the accuracy on
$\Delta m^2_{\rm atm}$ dramatically \cite{futureLBL}.  SNO will
install dedicated Neutral Current Detector (NCD) this fall, which will
allow event-by-event separation of CC/ES and NC events and lead to a
more accurate measurement of $\theta_{12}$ \cite{futureSNO}.  Later,
measurements of low-energy solar neutrino fluxes ($^7$Be and $pp$)
will allow even better determination of $\theta_{12}$
\cite{Bahcall:2003ce}.  The corrections due to $\theta_{13}$ will also
be constrained better by LBL experiments as well as new
multiple-baseline reactor anti-neutrino oscillation experiments
\cite{futurereactor}.

It is useful to recall the cosmological bound.  The combination of
WMAP, 2dFGRS, and Lyman $\alpha$ data leads to an upper bound
\cite{WMAP2} (see \cite{Hannestad:2003xv} for a slightly weaker bound)
\begin{equation}
  \sum_i m_{\nu_i} < 0.70~\mbox{eV},
\end{equation}
which translates to \cite{PM}
\begin{equation}
    |\langle m_\nu\rangle_{ee}| < 0.23~\mbox{eV},
\end{equation}
allowing the maximum constructive interference between three mass
eigenstates.  This follows from the fact that neutrinos are degenerate
in this mass range and the inequality
\begin{equation}
  |U_{e1}^2 + U_{e2}^2 + U_{e3}^2| \leq |U_{e1}^2| + |U_{e2}^2| + |U_{e3}|^2
  = 1.
\end{equation}
For a comparison \cite{PM}, the reported evidence for the neutrinoless
double beta decay suggest $|\langle m_\nu\rangle_{ee}|=
(0.11$--$0.56)$~eV \cite{Klapdor-Kleingrothaus}, while the reanalysis
in \cite{Vogel} gives 0.4--1.3~eV using a different set of nuclear
matrix elements.

To summarize, we have obtained a robust lower bound on the effective
electron-neutrino mass relevant to the neutrinoless double beta decay.
For the degenerate and inverted mass spectra, the next generation
experiments that have sensitivity on $|\langle m_\nu\rangle_{ee}|$
down to 0.01~eV can determine if neutrino is its own anti-particle.
For the normal hierarchy, the effective electron-neutrino mass may
even vanish.  However, if the large-scale structure cosmological data,
improved data on the tritium beta decay, or the long-baseline neutrino
oscillation experiments establish the degenerate or inverted mass
spectrum, the null result from such double-beta decay experiments will
lead to a definitive result.

\acknowledgments{This work was supported by Institute for Advanced
  Study, funds for Natural Sciences.  CPG is grateful to John Bahcall
  and Concha Gonzalez-Garcia for using partial results of the ongoing
  work on the analyses of solar and reactor data after SNO phase II
  data.  CPG was supported in part with the aid from the W.M.~Keck
  Foundation and in part by the NSF grant PHY-0070928.  HM was also
  supported in part by the DOE Contract DE-AC03-76SF00098 and in part
  by the NSF grant PHY-0098840.}

\end{document}